\documentclass[%
 reprint,
 amsmath,amssymb,
 aps,
]{revtex4-2}

\usepackage{graphicx}% Include figure files
\usepackage{dcolumn}% Align table columns on decimal point
\usepackage{bm}% bold math
\usepackage{textcomp}
\usepackage{hyperref}% add hypertext capabilities
\usepackage{url}
\usepackage{tkz-base,tkz-euclide}
\usepackage{caption}
\usepackage{subcaption}
\usepackage{graphicx}

\usepackage{comment}
\usepackage{xcolor}
\usepackage{float}

\begin{document}

\preprint{APS/123-QED}

%\title{Dynamics of Humans vs Zombies}
%\title{Dynamics of Humans Chased by Zombies}
\title{Simulating Pedestrian Avoidance: The Humans vs Zombies Scenario}

\author{Juan P. Oriana}
 \email{joriana@itba.edu.ar}
 \affiliation{%
 Instituto Tecnológico de Buenos Aires (ITBA), Lavardén 315, 1437 C. A. de Buenos Aires, Argentina
}%
\author{German A. Patterson}%
 \email{gpatters@itba.edu.ar}
\author{Daniel R. Parisi}%
 \email{dparisi@itba.edu.ar} 
\affiliation{%
 Instituto Tecnológico de Buenos Aires (ITBA), CONICET, Lavardén 315, 1437 C. A. de Buenos Aires, Argentina
}%

\date{\today}% It is always \today, today,

\begin{abstract}
This study introduces a unique active matter system as an application of the pedestrian collision avoidance paradigm, that proposes dynamically adjusting the desired velocity. We present a fictitious human-zombie scenario set within a closed geometry, combining prey-predator behavior with a one-way contagion process that can transform prey into predators. The system demonstrates varied responses: in cases where agents have the same maximum speeds, a single zombie always catches a human, whereas two zombies never catch a single human. As the number of human agents increases, observables, such as the final fraction of zombie agents and total conversion times, exhibit a significant change in the system's behavior at intermediate density values. Most notably, there is evidence of a first-order phase transition when the mean population speed is analyzed as an order parameter.

\end{abstract}

%Use showkeys class option if keyword
\maketitle

\section{Introduction}

A fundamental aspect in pedestrian dynamics modeling is describing how pedestrians navigate and avoid collisions in various environments. Initially, force-based models like the Social Force Model \cite{Helb1995, Helb2000} and the Predictive Collision Avoidance Model \cite{karamouzas2009predictive} simulate collision avoidance maneuvers by applying elusive forces directly on the agents.

An alternative framework, as proposed by Wang et al. \cite{wang2015new} and expanded in \cite{martin2020pedestrian, Martin2020, martin2020data, martin2023anisotropic}, involves dynamically adjusting the agents' desired velocity for navigation and avoidance maneuvers. This method closely resembles the natural walking behavior, wherein individuals adapt their speed and direction according to their surroundings, especially in scenarios where physical contact is absent. Moreover, this approach can be implemented in any microscopic agent model that has a specified target velocity, irrespective of whether the movement is defined by forces or rules.

As an alternative experimentation platform for this framework, we suggest exploring a fictional and basic system that imitates a predator-prey scenario. This model involves agents resembling zombies ($Z$) attacking humans ($H$), drawing inspiration from prevalent depictions in popular culture, including movies, series, comics, and novels. 

Other fictional systems involving zombies have previously been explored as an analogy for understanding the transmission of infectious diseases \cite{munz2009zombies, alemi2015you, bauer2019mathematical}. These studies often consider macroscopic population variables governed by differential equations to model their dynamics. On the other hand, adopting a microscopic approach, Libál et al. \cite{libal2023transition} investigate an $SCZR$ (Susceptible-Cleric-Zombie-Recovered) model by employing a population of self-propelled particles performing random motion.

In our proposed zombie-human system, particles exhibit movement directions determined by a specific heuristic. Zombie agents possess a desired velocity pointing toward the nearest human agent, while human agents aim to avoid zombies and collisions with walls and other humans. Additionally, our system is simplistic in terms of infection, following a $SI$ (Susceptible-Infected) model. The agents have only two states, and the only possible transition is unidirectional: $H \rightarrow Z$.

Beyond offering a distinctive and imaginative framework for the pedestrian navigation paradigm mentioned earlier, our proposed setting intriguingly amalgamates two concepts within a unified system: a) predator-prey dynamics and b) infectious disease contagion. This scenario stands out for its distinctive aspect: the contagion not only affects the prey but also transforms it into a predator.

The paper is structured as follows: Sec.~\ref{sec:model} introduces the microscopic model, while Sec.~\ref{sec:sim} outlines the simulation scenario. Sec.~\ref{sec:res} presents the outcomes of numerous simulations conducted to characterize the system. Finally, Sec.~\ref{sec:conc} contains our concluding remarks.

\section{The Human-Zombie Microscopic Model}
\label{sec:model}

The model considers two types of agents: zombies and humans. Common to both, the rule-based operational model used is the CPM \cite{Bagli2011}, plus a new avoidance layer that modifies the agent's desired velocity to prevent collisions. For human agents, the goal is to evade zombies, walls and other humans, while for zombies, the objective is to target the nearest human.

The original CPM assigns each agent a radius that can vary within the range from ${r_{min}}$ to $r_{max}$ following specific rules. These limiting values are shared by all agents in the simulation, regardless of whether they are zombies or humans.

The agent's radius is directly tied to the agent's speed $v_i$ as
\begin{equation}
    |\mathbf{v}_i| = v_i = v_i^{max}\left(\frac{r-r_{min}}{r_{max}-r_{min}}\right)^{\beta}\ ,
    \label{eq:1}
\end{equation}
where $\beta$ is a constant.
The agent's maximum velocity $v^{max}$ is the free velocity and it is distinctive for each agent category:
\begin{equation}
    v_i^{max} =
    \begin{cases}
    v_z^{max} & \text{if agent $i$ is a zombie}\ , \\
    v_h^{max} & \text{if agent $i$ is a human}\ .
    \end{cases}
    \label{eq:2}
\end{equation}
The velocity of each agent is adjusted in each discrete temporal increments $\Delta t$ by Eq.~\eqref{eq:1}. Simultaneously, the radius and position of each agent are updated according to:
\begin{align}
    \label{eq:3} r_i(t + \Delta{t}) & = \min{\left[r_{max}, r_i(t) + \frac{r_{max}}{\tau/\Delta{t}}\right]}\ , \\
    \mathbf{x}_i(t+\Delta{t}) & = \mathbf{x}_i(t) + \mathbf{v}_i \Delta{t}\ ,
    \label{eq:4}
\end{align}
where $\tau$ is a characteristic time constant describing the duration it takes for a particle to reach its maximum speed after starting from rest.

On the other hand, when a collision occurs, the state of the involved particles is updated as follows. First, they contract their radii to the minimal value, $r_{min}$. Then, at the next time step, the agents undergo repulsion using an escape velocity, $\textbf{v}^e$, in a direction determined by the center of two particles, each moving away from each other. The magnitude of this escape velocity corresponds to the maximum velocity, $v^e = v^{max}$.

Moreover, in our specific scenario, the interaction between humans and zombies results in the immobilization of both agents for a period $t_c$, representing the contagion process. Upon completion, the human agent transforms into a zombie and emerges with a radius of $r = r_{min}$.

Additionally, to counteract artifacts arising from symmetric configurations, we add a small angular noise to the direction of the desired velocity. Let $\theta_i$ represent the angle formed by $\textbf{v}_i$ with respect to the positive x-semiaxis, and let $\theta^t_i$ denote the angle of the desired direction vector of agent $i$, thus, $\theta_i = \theta_i^t + \Delta \theta$, with $\Delta \theta$ defined as a uniform random value within the range of $[-\mu/2, \mu/2]$.

In what follows, we describe the heuristic that allows the avoidance layer to dynamically adjust the direction of the desired velocity of agents: $\mathbf{v}_i = v_i \mathbf{e}_i^t$. This direction is represented by the unit vector $\mathbf{e}^t_i = \frac{\mathbf{x}^t_i - \mathbf{x}_i}{{| \mathbf{x}^t_i -\mathbf{x}_i |}}$, which is determined by a temporal target denoted as $\mathbf{x}_i^t$. This target is computed differently depending on whether agent $i$ is a human or a zombie.

If agent $i$ is a human, $\mathbf{{x}}_i^t$ is obtain from the sum of the collision vectors $\mathbf{n}_c^{ij}$. Each $\mathbf{n}_c^{ij}$ quantifies the extent to which a human $i$ should adjust its direction to move away from an agent $j$. Therefore, this desired target is defined by Eq. \ref{eq:5}.
\begin{equation}
    \mathbf{x}_i^t = \sum_{z \in Z^i}\mathbf{n}_c^{iz} +
    \sum_{h \in H^i} \mathbf{n}_c^{ih} +
    \mathbf{n}_c^{iw} \ ,
    \label{eq:5}
\end{equation}
where $Z^i$ represents the set of the $N^e_z$ nearest zombies to agent $i$, $H^i$ the set of the $N^e_h$ nearest humans to agent $i$. 

The vectors $\mathbf{n}_c^{ij}$ are computed as:
\begin{equation}
    \mathbf{n}_c^{ij} = \mathbf{e}^{ij} A e^{-\frac{d_{ij}}{B}} \ ,
    \label{eq:6}
\end{equation}
where $d_{ij}$ is the distance between agents, $\mathbf{e}^{ij} = \frac{\mathbf{x}^i - \mathbf{x}^j}{|\mathbf{x}^i - \mathbf{x}^j|}$ is the direction from agent $i$ to agent $j$, and $A$ and $B$ are constants denoting the intensity and a characteristic length, respectively. The latter two parameters vary depending on whether agent $j$ is a human, a zombie, or the wall itself.

The last term, $\mathbf{n}_c^{iw}$, accounts for the interaction of agent $i$ with walls ($j=w$). In this specific case, we will consider $\mathbf{x}^w$ as the nearest point on the boundary to agent $i$.

If agent $i$ is a zombie, its target is the nearest human agent $h$, defined by $\mathbf{x}_i^t = \mathbf{e}^{ih}$.

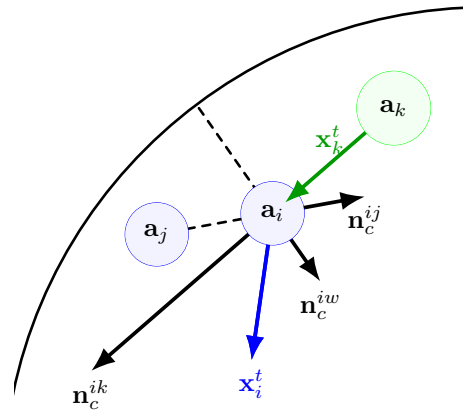
\begin{figure}[!h]
\centering
    \begin{tikzpicture}[thick, scale=0.7]
    \tkzDefPoints{0.9/0.3/H1, 1.5/0.3/h1, -1.3/-0.1/H2, -1.3/-0.7/h2, 3.2/2.3/Z, 3.9/2.3/z, 5/-5/R, -1/2/r,
    -4/-4.0/A, 4.5/-4.0/B, 4.5/4.5/C, -4/4.5/D, -0.5/2.3/W,
    0.5/-2.5/T}

    \tkzDefPointBy[homothety=center Z ratio .9](H1)
    \tkzGetPoint{toH}
    \tkzDefPointBy[homothety=center H1 ratio .85](Z)
    \tkzGetPoint{toZ}

    \tkzDrawSegments[style=dashed, line width =1](H1,H2)
    \tkzDrawSegments[style=dashed, line width =1](Z,H1)
    \tkzDrawSegments[style=dashed, line width =1](W,H1)

    \tkzDefPointOnLine[pos=1.8](H2,H1)\tkzGetPoint{EH}
    \tkzDefPointOnLine[pos=2.5](Z,H1)\tkzGetPoint{EZ}
    \tkzDefPointOnLine[pos=1.65](W,H1)\tkzGetPoint{EW}

    \tkzDrawSegments[vector style,line width =1.4](H1,EH)
    \tkzDrawSegments[vector style,line width =1.4](H1,EZ)
    \tkzDrawSegments[vector style,line width =1.4](H1,EW)
    \tkzDrawSegments[vector style,line width =1.4,color=blue](H1,T)

    \tkzDrawSegments[vector style,line width =1.4,color=blue](H1,T)

    \tkzDrawCircle[blue](H1,h1)
    \tkzFillCircle[white](H1,h1)
    \tkzFillCircle[opacity=0.05,blue](H1,h1)
    \tkzDrawCircle[blue](H2,h2)
    \tkzFillCircle[white](H2,h2)
    \tkzFillCircle[opacity=0.05,blue](H2,h2)

    \tkzDrawSegments[vector style,line width =1.4,color=green!60!black](toZ,toH)

    \tkzDrawCircle[green](Z,z)
    \tkzFillCircle[white](Z,z)
    \tkzFillCircle[opacity=0.05,green](Z,z)

    \tkzClipPolygon(A,B,C,D)
    \tkzDrawCircle[black,line width=1](R,r)
 
    \tkzLabelPoint[anchor=center](Z){$\mathbf{a}_k$}
    \tkzLabelPoint[anchor=center](H1){$\mathbf{a}_i$}
    \tkzLabelPoint[anchor=center](H2){$\mathbf{a}_j$}

    \tkzLabelPoint(EH){$\mathbf{n}_c^{ij}$}
    \tkzLabelPoint(EZ){$\mathbf{n}_c^{ik}$}
    \tkzLabelPoint(EW){$\mathbf{n}_c^{iw}$}
    \tkzLabelPoint[blue](T){$\mathbf{x}_i^t$}
    \tkzLabelSegment[green!60!black, above](toZ,toH){$\mathbf{x}_k^t$}

    \end{tikzpicture}
    
\caption{Selection of the dynamic target $\textbf{x}_i^t$ for a human agent $a_i$ depends on the positions of the wall, another human $a_j$ (blue) and a zombie agent $a_k$ (green). Also depicted, selection of dynamic target $\textbf{x}_k^t$ for a zombie agent $a_k$, solely dependent on human agent $a_i$.}
    \label{Avoidance}
\end{figure}

All the previously defined vectors can be visualized in Fig. \ref{Avoidance}

\section{Simulations}
\label{sec:sim}

\begin{figure}[!h]
\centering
    \begin{tikzpicture}[thick, scale=0.3]
    \tkzInit[xmin=0,xmax=1.5,ymin=-0,ymax=1.5]
    \tkzDefPoints{0/0/Z, 0/0.35/z, 11/0/Ra, 1/0/Ri, 10/0/Ro, 5/7/Ha, 5.35/7/ha, -3/-4/Hb, -3.35/-4/hb, 7/-5/Hc, 7.35/-5/hc, -5/4/Hd, -5/4.35/hd, 0.1/4.8/He, 0.1/5.15/he}
    
    \tkzDrawCircle[black](Z,Ra)

    \tkzFillCircle[opacity=0.05,red](Z,Ra)
    \tkzFillCircle[white](Z,Ro)
    \tkzFillCircle[opacity=0.05,red](Z,Ri)

    \tkzDrawCircle[black, style=dashed, line width = 0.6](Z,Ri)
    \tkzDrawCircle[black, style=dashed, black, style=dashed, line width = 0.6](Z,Ro)
    
    \tkzDrawCircle[black,line width=1](Z,z)
    \tkzFillCircle[white](Z,z)
    \tkzDrawCircle[black](Ha,ha)
    \tkzFillCircle[opacity=0.15,blue](Ha,ha)
    \tkzDrawCircle[black](Hb,hb)
    \tkzFillCircle[opacity=0.15,blue](Hb,hb)
    \tkzDrawCircle[black](Hc,hc)
    \tkzFillCircle[opacity=0.15,blue](Hc,hc)
    \tkzDrawCircle[black](Hd,hd)
    \tkzFillCircle[opacity=0.15,blue](Hd,hd)
    \tkzDrawCircle[black](He,he)
    \tkzFillCircle[opacity=0.1,blue](He,he)

    \tkzDrawSegments[style=dashed, line width =0.5](Z,Ra)
    \tkzLabelSegment[black](Z,Ra){$R_a$}
    \tkzDrawSegments[vector style, line width = 0.5](Z,Hd)
    \tkzLabelSegment[black](Z,Hd){$\mathbf{x}_i$}

    \tkzFillCircle[white](Z,z)
    \tkzFillCircle[opacity=0.15,green](Z,z)

    \tkzDrawXY

    \end{tikzpicture}
\caption{Illustration of the simulation domain and the initial configuration featuring a zombie at the center (green agent) and $N_h=5$ human agents (blue agents).}
\label{figArena}
\end{figure}
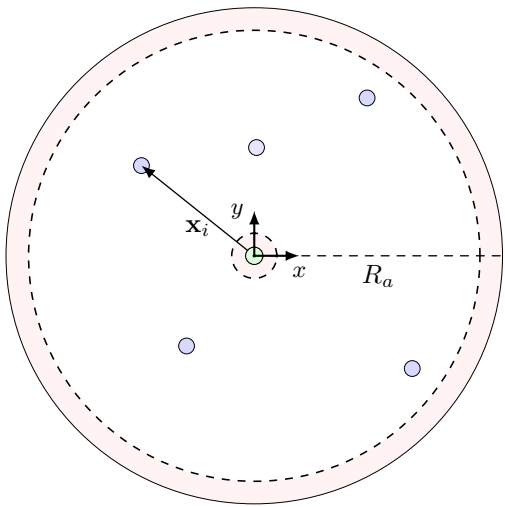

The agents move within a circular arena with a radius $R_a = 11$ m, as shown in Fig. \ref{figArena}. The simulations start with a solitary zombie agent positioned at the center of the domain, and $N_h(t=0) \equiv N_h$ human agents randomly distributed, ensuring that their distance from both the center and the wall exceeds 1 m (1 m $ \leq |\mathbf{x}_i| \leq R_{a} -$ 1 m), as illustrated by the dotted circles in Fig.~\ref{figArena}. 

Considering the model described in Sec.~\ref{sec:model}, the fixed parameters for the agents are the following: $\Delta t = 0.05$ s; $r_{min} = 0.15$ m; $r_{max} = 0.35$ m; $v_h^{max} = 4.0$ m/s; $\beta = 0.9$; $\tau = 0.5$ s; $\mu = 0.052$ rad (3\textdegree); $A_h = 4$; $B_h = 1$ m; $N^e_h = 1$; $A_z = 8$; $B_z = 4$ m; $N^e_z = 2$; $A_w = 8$; $B_w = 1$ m; $t_c = 3$ s.

The initial number of humans ($N_h$) will vary between 10 and 100, in steps of 5, and is considered the control parameter. It is important to note that the total number of agents is $N=N_h+1$ ($N_h$ humans plus one zombie), and this number remains constant in each simulation as human agents turn into zombies. Moreover, the free speed of the zombie, $v_z^{max}$, is swept from 3.8 m/s to 4.2 m/s in steps of 0.1 m/s, being the reference case $v_z^{max}=4$ m/s . Overall, there are a total of 95 system configurations to explore. The statistical analysis of each was conducted on 2500 realizations of random initial conditions. Each realization continued until either all agents had become zombies or the simulated time reached $t_{max} = 2000$ s. We define the final simulation time as $T$, and the number of simulation steps as $S$. Hence, $S = \frac{T}{\Delta{t}}$. In the case that $T = t_{max}$, then $S=40,000$.

We will now define the observables under investigation. First, the fraction of zombies at any given time $t$ in a realization $k$ is given by:
\begin{equation}
\phi^k_z(t) = \frac{N^k_z(t)}{N^k_h + 1}= \frac{N^k_z(t)}{N}
\label{eq:8}
\end{equation}
and compute $\left\langle\phi_z(t)\right\rangle$, which represents the average of $\phi_z^k$ across the realizations. We focus our study on the endpoint of the average final zombie fraction (hereafter referred to simply as the final zombie fraction), denoted as $\left\langle\phi_z^{final}\right\rangle$, which is defined when $\left\langle \phi_z(t)\right\rangle$ reaches a stationary value.

The second observable we calculate is the time required to achieve total conversion, denoted as the 'total conversion time', $T_c$, and we average it across all realizations as the 'average total conversion time', $\left\langle T_c\right\rangle$. Any realization that did not achieve total conversion within $t_{max}$ was excluded from this particular analysis. If fewer than 10\% of realizations achieve total conversion, that specific configuration will not be considered for analysis.

The last observable we define is the mean velocity, used as a type of order parameter. This is based on the understanding that zombie-human interactions typically occur at zero velocity, consequently reducing the mean velocity of the agents. This reduction is even more significant than the usual decrease caused by density in pedestrian dynamics and the CPM model \cite{Bagli2011}.

Considering a specific realization $k$ at time $t$, the system's velocity, accounting for both agent types, is expressed as follows:
\begin{equation}
v^k(t) = \frac{1}{N} \sum_{i=1}^{N} |\mathbf{v}_i^k(t)|\ ,
\label{eq:9}
\end{equation}
where $\mathbf{v}_i$ is the vector velocity of particle $i$. Then, we compute the average across the realizations as:
\begin{equation}
\bar{v}= \bar{v}(t) = \frac{1}{K} \sum_{k=1}^{K} v^k(t)\ ,
\label{eq:10}
\end{equation}
with $K=2500$, the number of realizations per configuration.
Lastly, in the case of stationary states, the time average of mean population velocity is computed as $\left\langle\bar{v}\right\rangle_t$.

\section{Results}
\label{sec:res}

\subsection{Dynamics involving 2 and 3 agents}
\label{Sec:fewHumans}
The behaviors of the system when $N_h=1$ and $N_h=2$ are interesting and notably different from one another. In the simplest scenario, the system consists of a single human agent ($N_h=1$) in addition to the initial zombie ($N_z(t=0)=1$) within the simulation domain. We will show that, under conditions of equal maximum speed ($v_z^{max} = v_h^{max} = v^{max} $), the zombie will always catch the human agent.

\begin{figure}
    \centering
    \includegraphics[width=\columnwidth]{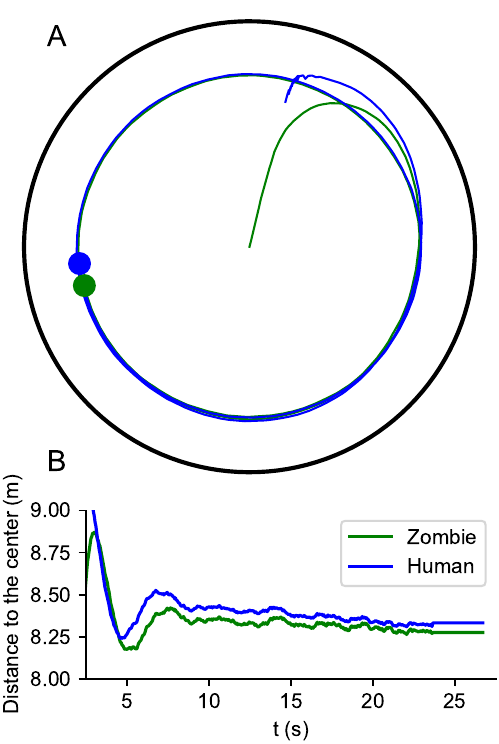}
    \caption{Dynamics with 2 agents. A: Human agent evading a zombie while moving along the curved wall. B: Distance of agents from the center of the enclosure.}
    \label{TwoAgents}
\end{figure}

In the long-term pursuit between a human agent and a zombie, the human's trajectory becomes circular, running parallel to the enclosure's wall. This behavior is illustrated in Fig.~\ref{TwoAgents}A, showing the trajectories of both agents. Fig.~\ref{TwoAgents}B depicts the distance of both agents from the center of the enclosure over time. The zombie agent pursues the prey with a smaller turning radius, resulting in a higher angular velocity. Consequently, the zombie always catches up with its prey.  See Supplemental Material \cite{SMV1} for a video showing this dynamics.

This outcome indicates that in our simulations, when the speed of zombies is similar to that of humans, the initial zombie will inevitably succeed in converting at least one human. This claim is strengthened by the fact that human agents must evade not just walls, but also other agents, which, on top of that, can reduce their speed.

An important consideration is the potential for a human to escape if a model incorporating inertia and the capability for zig-zag evasive maneuvers is employed. Under these conditions, the aforementioned conclusion regarding an isolated human escaping from a zombie would no longer hold. 

Let us now examine the immediately more intricate case ($N_h=2$), showcasing behavior markedly distinct from the $N_h=1$ scenario. Under this condition, the zombie chases the nearest human until it catches him, which will definitely happen, as demonstrated above. After converting the human into a zombie, the system's state changes to $N_z(t)=2$ and $N_h(t)=1$. Given these conditions, with both zombies having only one remaining human target, they both move towards that individual. With the two zombies being in close proximity following the conversion, they collide while moving towards the same target. This collision decelerate the zombies, granting the human an opportunity to accelerate away from both of them at maximum speed. As they resume their approach towards the target, after a few time steps, they collide once more, again diminishing their speed and allowing the human to distance themselves again. This pattern repeats indefinitely, ensuring that two zombies, with maximum desired speeds equal to that of the human, can never catch him, within the context of this operational model. An animation of this dynamics can be seen on Supplemental Material \cite{SMV2}.

Summing up, under the given conditions, when $N_h=1$ the conversion will always be complete, i.e.: $\phi_z=1$. In contrast, this will never occur when $N_h=2$, resulting in a stable zombie fraction of $\phi_z=2/3$.

\subsection{Dynamics involving more than 3 agents}

Having previously analyzed the fundamental cases of one and two humans, we now turn our attention to more densely populated initial conditions.

\begin{figure}[ht]
    \centering
    \begin{minipage}{0.15\textwidth}
        \centering
        \begin{tikzpicture}
             \draw (0, 0) node[inner sep=0] {\includegraphics[width=\linewidth]{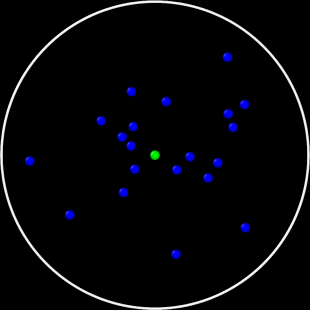}};
            \draw (-1.15,1.15) node[text=white, font=\sffamily] {A};
         \end{tikzpicture}
    \end{minipage}%
    \begin{minipage}{0.15\textwidth}
        \centering
        \begin{tikzpicture}
             \draw (0, 0) node[inner sep=0] {\includegraphics[width=\linewidth]{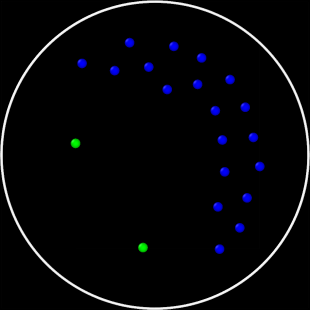}};
            \draw (-1.15,1.15) node[text=white, font=\sffamily] {B};
         \end{tikzpicture}
    \end{minipage}%
    \begin{minipage}{0.15\textwidth}
        \centering
        \begin{tikzpicture}
             \draw (0, 0) node[inner sep=0] {\includegraphics[width=\linewidth]{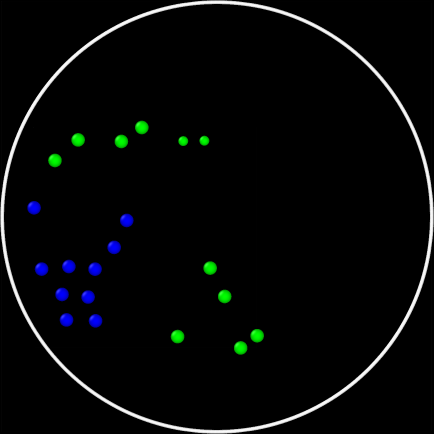}};
            \draw (-1.15,1.15) node[text=white, font=\sffamily] {C};
         \end{tikzpicture}
    \end{minipage}

    \begin{minipage}{0.15\textwidth}
        \centering
        \begin{tikzpicture}
             \draw (0, 0) node[inner sep=0] {\includegraphics[width=\linewidth]{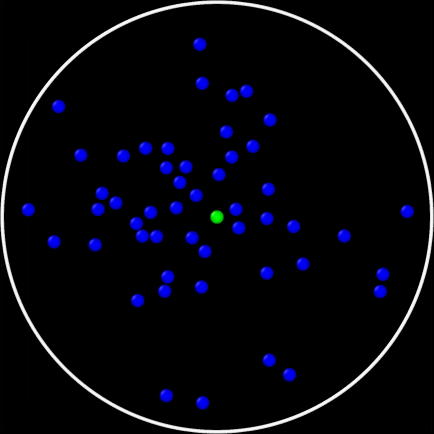}};
            \draw (-1.15,1.15) node[text=white, font=\sffamily] {D};
         \end{tikzpicture}
    \end{minipage}%
    \begin{minipage}{0.15\textwidth}
        \centering
        \begin{tikzpicture}
             \draw (0, 0) node[inner sep=0] {\includegraphics[width=\linewidth]{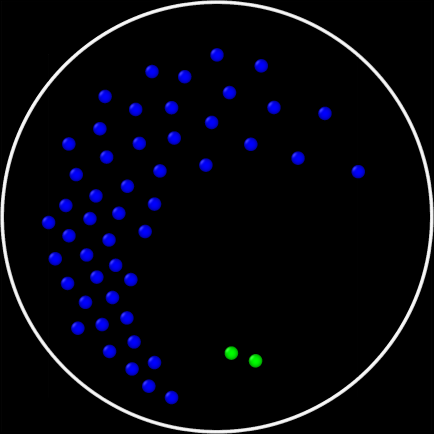}};
            \draw (-1.15,1.15) node[text=white, font=\sffamily] {E};
         \end{tikzpicture}
    \end{minipage}%
    \begin{minipage}{0.15\textwidth}
        \centering
        \begin{tikzpicture}
             \draw (0, 0) node[inner sep=0] {\includegraphics[width=\linewidth]{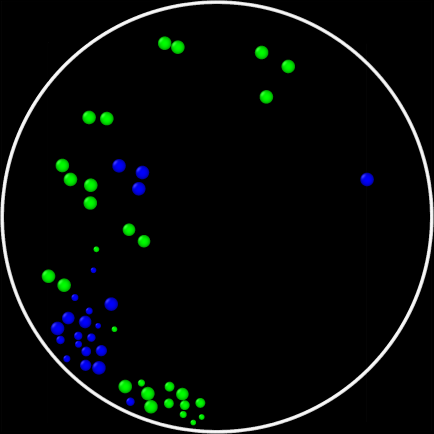}};
            \draw (-1.15,1.15) node[text=white, font=\sffamily] {F};
         \end{tikzpicture}
    \end{minipage}

    \begin{minipage}{0.15\textwidth}
        \centering
        \begin{tikzpicture}
             \draw (0, 0) node[inner sep=0] {\includegraphics[width=\linewidth]{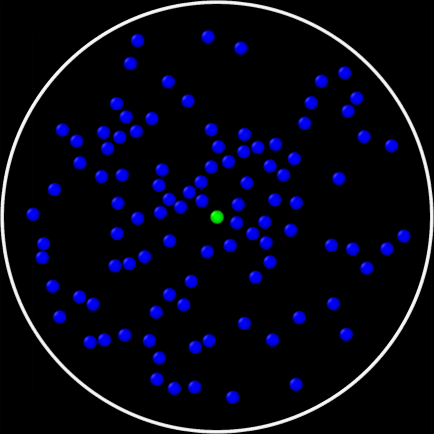}};
            \draw (-1.15,1.15) node[text=white, font=\sffamily] {G};
         \end{tikzpicture}
    \end{minipage}%
    \begin{minipage}{0.15\textwidth}
        \centering
        \begin{tikzpicture}
             \draw (0, 0) node[inner sep=0] {\includegraphics[width=\linewidth]{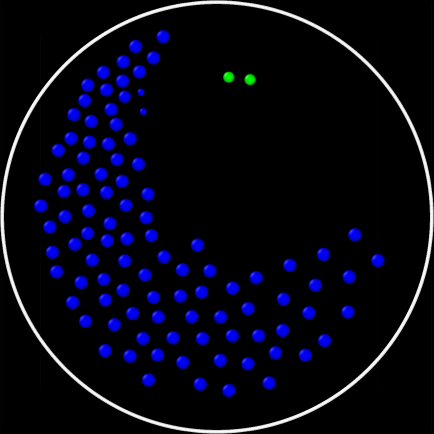}};
            \draw (-1.15,1.15) node[text=white, font=\sffamily] {H};
         \end{tikzpicture}
    \end{minipage}%
    \begin{minipage}{0.15\textwidth}
        \centering
        \begin{tikzpicture}
             \draw (0, 0) node[inner sep=0] {\includegraphics[width=\linewidth]{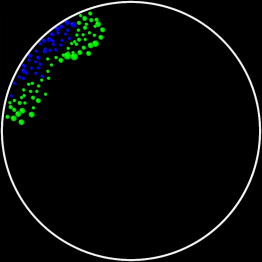}};
            \draw (-1.15,1.15) node[text=white, font=\sffamily] {I};
         \end{tikzpicture}
    \end{minipage}
    \caption{Temporal evolution of three exemplary configurations. A: $N_h = 20$ at $t=0$. B: $N_h = 20$ at first conversion. C: $N_h = 20$ at $\phi_z=0.5$. D: $N_h = 50$ at $t=0$. E: $N_h = 50$ at first conversion. F: $N_h = 100$ at $\phi_z=0.5$. G: $N_h = 100$ at $t=0$. H: $N_h = 100$ at first conversion. I: $N_h = 100$ at $\phi_z=0.5$.}
    \label{Fig:Arena_Nh}
\end{figure}
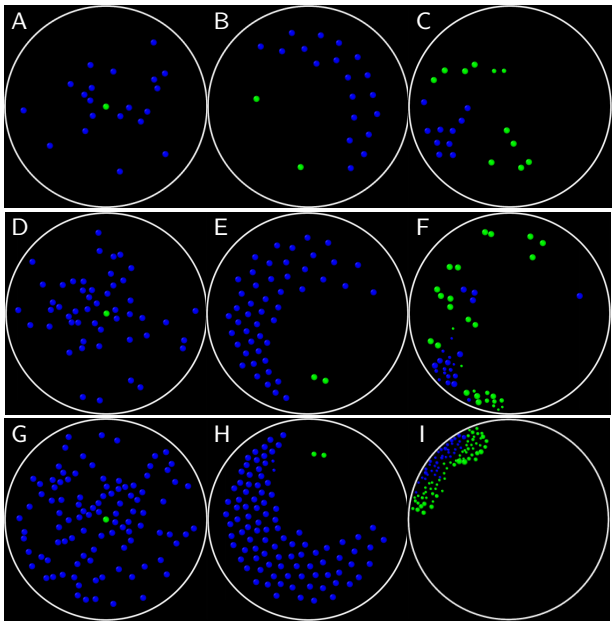

To illustrate the system's dynamics, we present several representative snapshots in Fig. \ref{Fig:Arena_Nh} for different initial values of $N_h$, taken at specific moments corresponding to distinct zombie fractions.

It is crucial to highlight that after the first conversion (when reaching $N_z=2$), the two zombies may either target the same human agent or different ones, depending on the positions of other human agents. In the former scenario, a loop could arise where the remaining humans can escape from the two zombies that continuously collide with each other (similar to the $N_h=2$ scenario discussed in the previous Sec. \ref{Sec:fewHumans}). This scenario is termed 'convergent pursuit,' and a video example is available in the Supplemental Material \cite{SMV3}. In the latter scenario, where the two zombies split to pursue different human agents, they will succeed, mirroring the $N_h=1$ situation described in Sec. \ref{Sec:fewHumans}. This is referred to as 'divergent pursuit,' and a video is available in the Supplemental Material \cite{SMV4}. Additionally, snapshots of both types of pursuit can be seen in Fig. \ref{Fig:Arena_Nh}B for 'divergent' and Fig. \ref{Fig:Arena_Nh}E and H for 'convergent'.

After establishing these definitions, simulations are categorized based on the following criteria: if, post the initial conversion, both zombies remain within a distance of less than 2 meters for 3 seconds, the scenario is classified as convergent pursuit. Otherwise, it falls under divergent pursuit. 

\begin{figure}
    \centering
     \includegraphics[width=\linewidth]{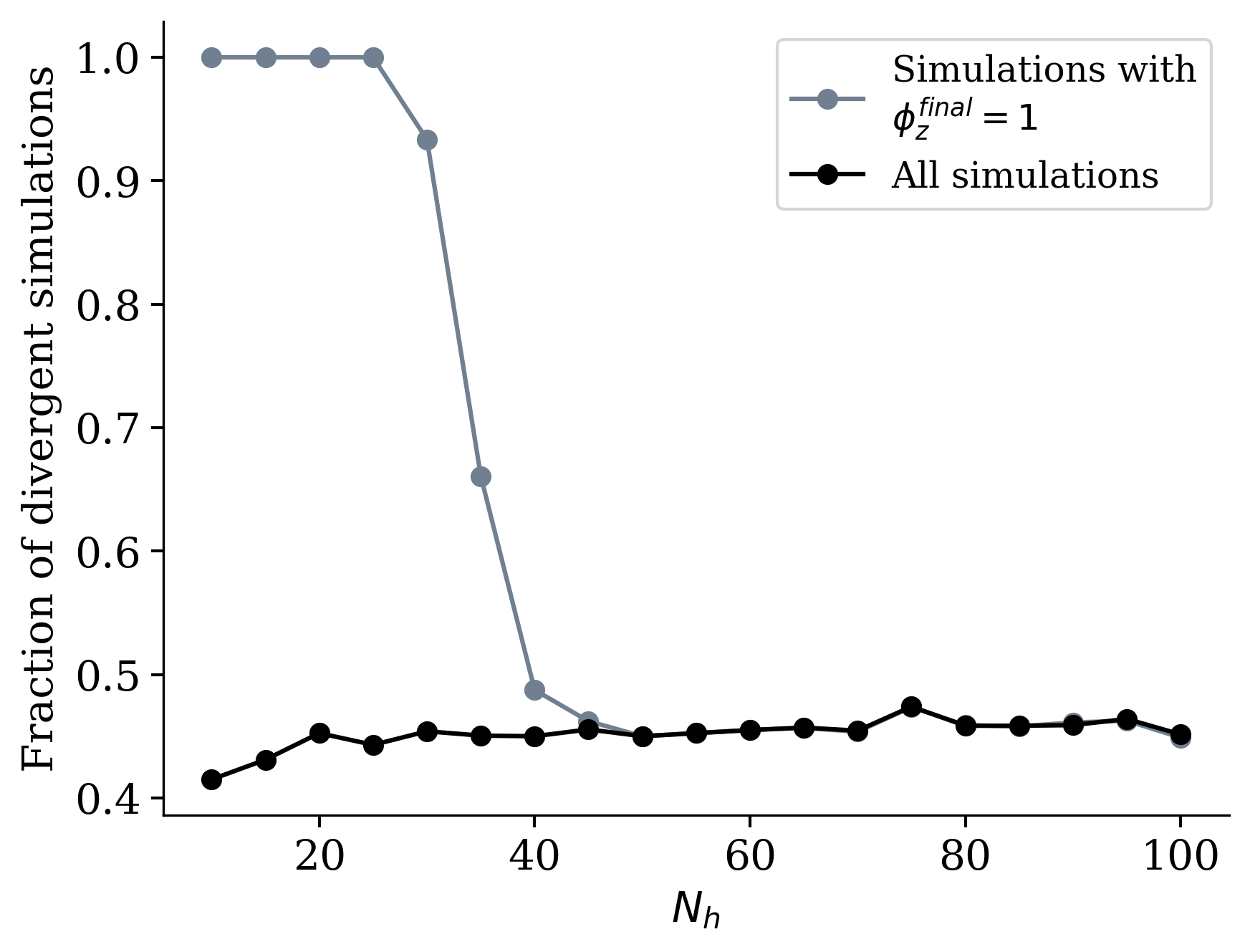}
    \caption{Fraction of divergent pursuit cases as a function of the number of initial human agents in the system with a maximum zombie speed of $v_z^{max}=v_h^{max}=4$ m/s.}
    \label{ConvergenceRatio}
\end{figure}

Figure~\ref{ConvergenceRatio} shows that the distribution of total simulations between each category is fairly balanced. However, when considering only simulations that result in total conversion, the only mechanism leading to total conversion for $N_h \leq 25$ is the divergent pursuit type. As $N_h$ increases, a transition is observed, reaching a balance after $N_h \geq 45$. In this region of elevated densities, both types of initial pursuits exhibit comparable probabilities of achieving total conversion due to the decelerating effect caused by the increased density of human agents.

\subsubsection{Final Zombie Fraction}

In this section, we examine the dynamics of the population, focusing on the average final zombie fraction, $\left\langle\phi_z^{final}\right\rangle$ [Eq.~\ref{eq:8}]. Since the number of agents remains constant in each simulation, the fraction of human agents is complementary to the zombie fraction: \\ $\left\langle\phi_z^{final}\right\rangle = 1-\left\langle\phi_h^{final}\right\rangle$. Consequently, we focus our discussion solely on the zombie fraction.

\begin{figure}
    \begin{subfigure}{\columnwidth}
        \centering
        \begin{tikzpicture}
             \draw (0, 0) node[inner sep=0] {\includegraphics[width=\linewidth]{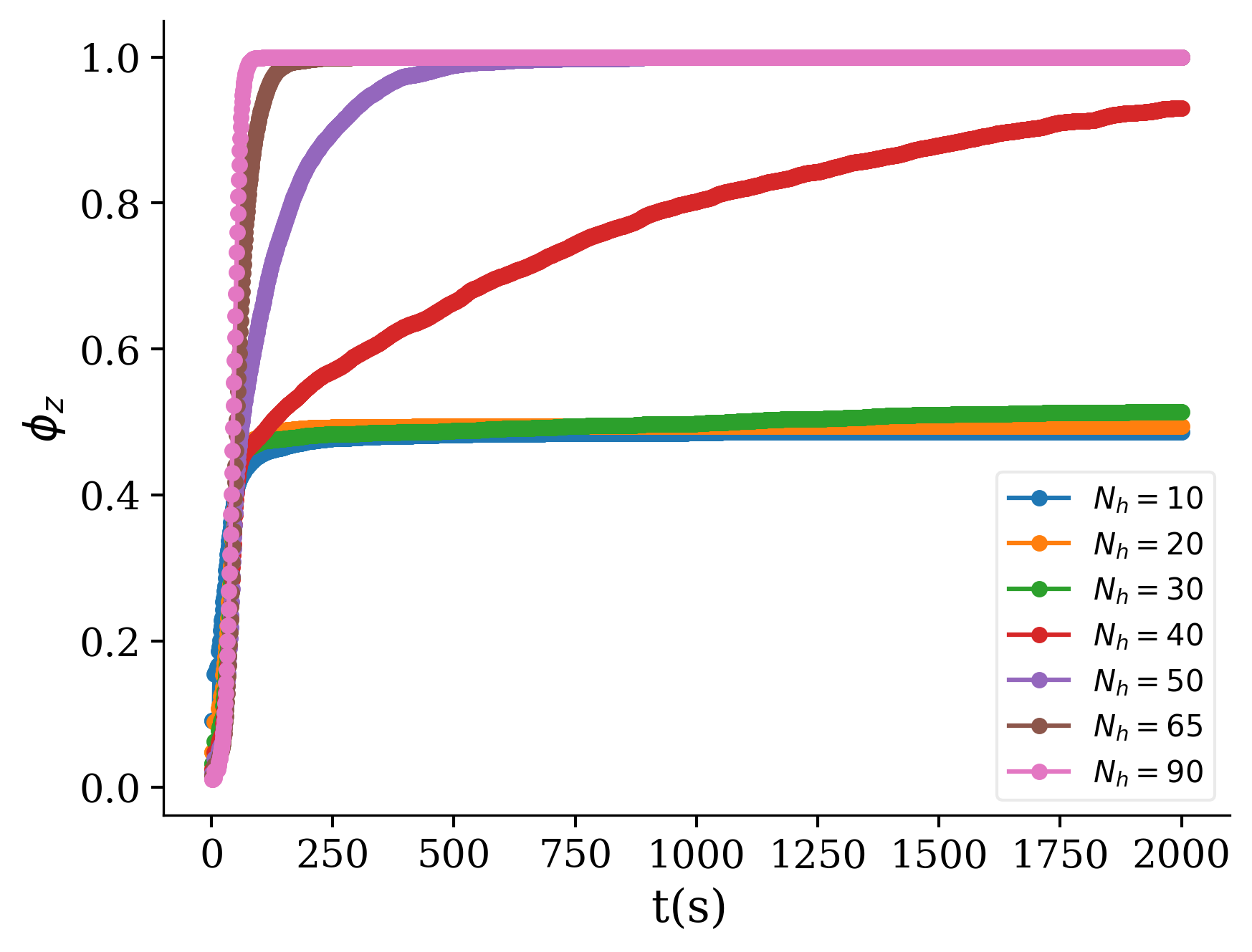}};
            \draw (-4,3.0) node[text=black, font=\sffamily] {\large A};
         \end{tikzpicture}
    \end{subfigure}
    % \vspace{10pt} % Adjust the vertical space between subfigures
    \begin{subfigure}{\columnwidth}
        \centering
        \begin{tikzpicture}
             \draw (0, 0) node[inner sep=0] {\includegraphics[width=\linewidth]{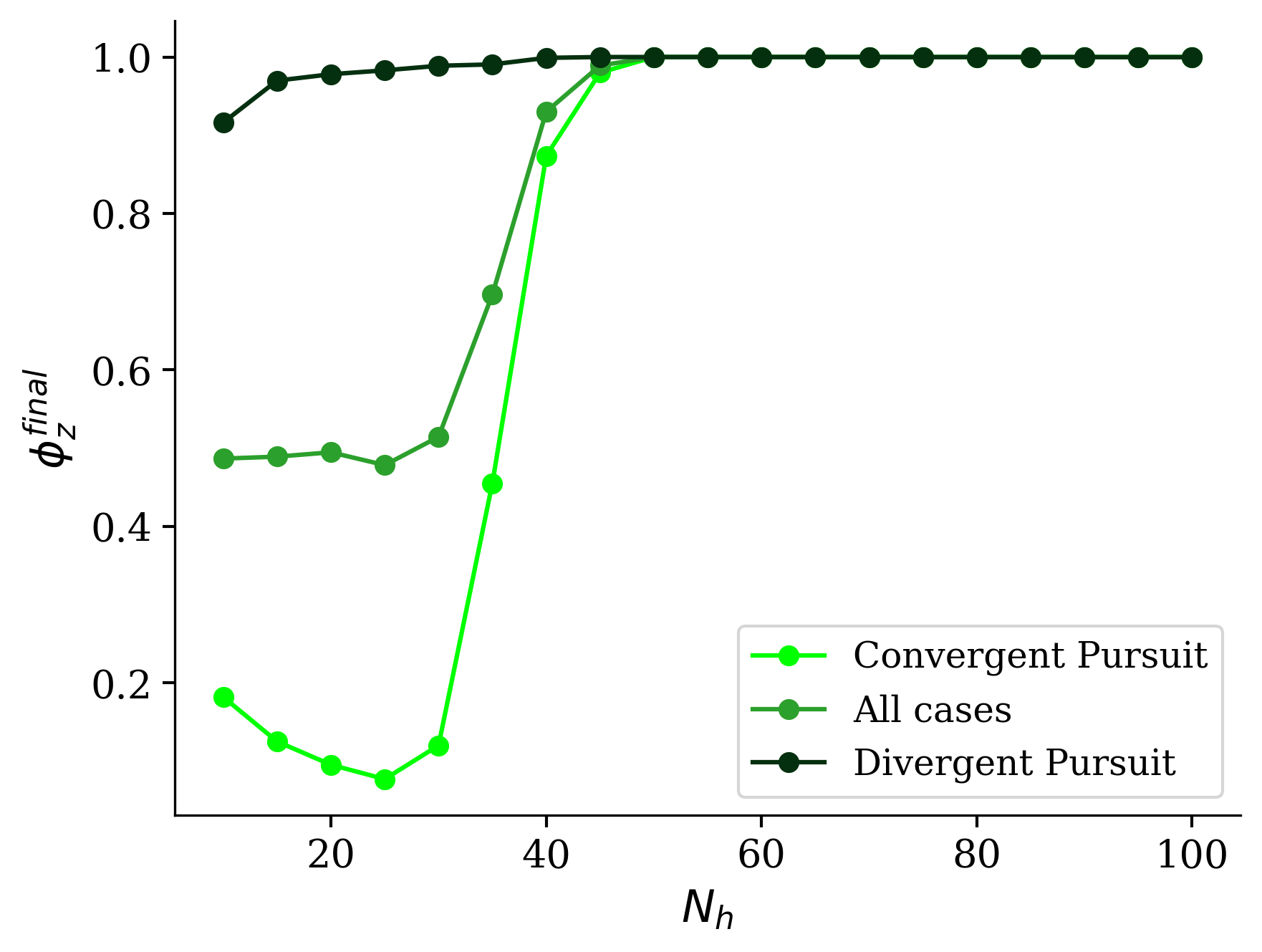}};
            \draw (-4,3.0) node[text=black, font=\sffamily] {\large B};
         \end{tikzpicture}
    \end{subfigure}
    \caption{Zombie Fraction. A: Time evolution of $\phi_z$ for $v_z^{max}=4$ m/s and the specified $N_h$ values. B: Final zombie fraction $\phi_z^{final}$ as a function of the initial number of human agents, categorized by the type of pursuit.}
    \label{fig:frac_per_time}
\end{figure}

First, we examine the time evolution of $\left\langle\phi_z(t)\right\rangle$ in Fig.~\ref{fig:frac_per_time}A for the scenario where $v_z^{max}=v_h^{max}=4$ m/s. The mean zombie fraction is observed to reach stationary values within the simulation time, clustering around low and high values of $N_h$. A notable transition occurs around intermediate values of $N_h \sim 40$, requiring more time to reach the stationary regime.

This transition becomes clearer in Fig~.\ref{fig:frac_per_time}A, where $\left\langle\phi_z^{final}\right\rangle$ is plotted against the initial number of human agents. The curve labeled 'All cases' represents the stationary data points from Fig.~\ref{fig:frac_per_time}A. For values up to $N_h=30$, the final zombie fraction is $\left\langle\phi_z^{final}\right\rangle \approx 0.5$. However, beyond $N_h=45$, the system invariably results in total conversion.

The intermediate values of $\left\langle\phi_z^{final}\right\rangle$ at lower $N_h$ arise from two extreme possible outcomes of the simulations: one with a very high zombie fraction and the other with a very low one. In the latter scenario, humans have ample space to outpace the zombies, who collide with each other, impeding their pursuit (convergent pursuit). Figure~\ref{fig:frac_per_time}B illustrates that these two extremes correspond to divergent and convergent pursuit scenarios. This confirms our earlier observations from the analysis of Fig. \ref{ConvergenceRatio}: total conversion at lower $N_h$ values is solely observed in cases of divergent pursuit.

On the other hand, when the initial human population exceeds $N_h\sim 45$, the constrained space inhibits human escape without mutual interference. This heightened density reduces their speed, invariably leading to zombies catching them, regardless of pursuit type.

\begin{figure}%[!ht]
    \centering
        \begin{tikzpicture}
             \draw (0, 0) node[inner sep=0] {\includegraphics[width=\linewidth]{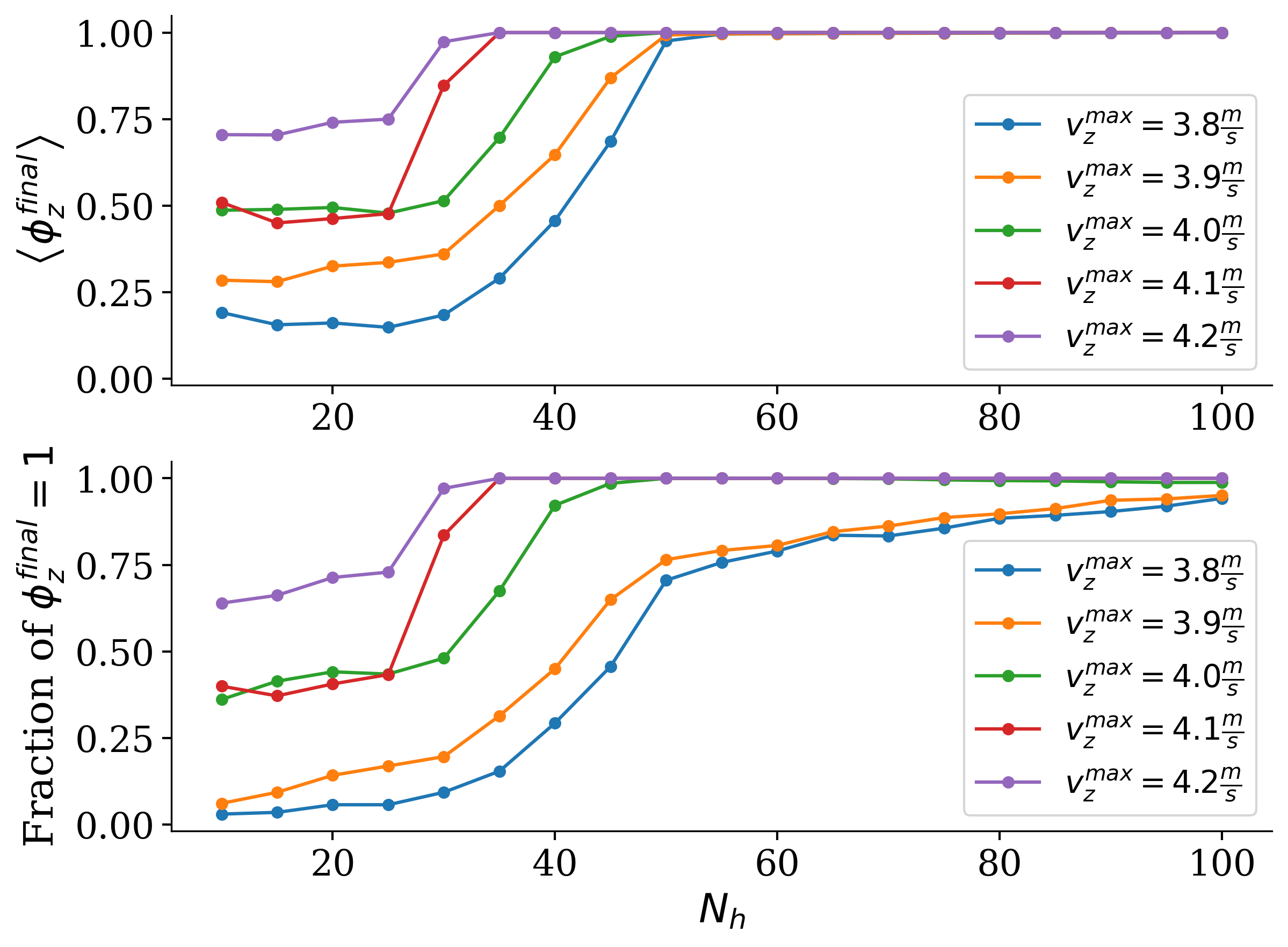}};
            \draw (-2.9,3.0) node[text=black, font=\sffamily] {\large A};
            \draw (-2.9,0.0) node[text=black, font=\sffamily] {\large B};
         \end{tikzpicture}
    \caption{Final Conversion for $v_z^{max}$ Lower and Higher than $v_h^{max}$. A: Final zombie fraction as a function of $N_h$ for various $v_z^{max}$ values. B: Ratio of realizations resulting in total conversion to the total number of simulations conducted for each $N_h$ and $v_z^{max}$ value.}
    \label{fig:success_rate}
\end{figure}

Now, we explore the same observable for varying maximum zombie speeds, maintaining a constant human maximum speed of $v_h^{max}=4$ m/s. Figure~\ref{fig:success_rate}A displays five curves corresponding to $v_z^{max} \in [3.8, 4.2]$ m/s. Observations reveal that the curves qualitatively exhibit a similar pattern, showing a transition from partial to total conversion at intermediate $N_h$ values. However, as $v_z^{max}$ increases, the final zombie fraction at lower $N_h$ values decreases, and the $N_h$ value at which the system saturates becomes higher. Both observations support the concept that higher-speed zombies can capture all human agents at lower critical densities. Additionally, the likelihood of converting the entire population becomes higher at these lower densities.

In Fig.\ref{fig:success_rate}B the fraction of simulations achieving total conversion is shown. For the cases of $v_z^{max} \geq v_h^{max} = 4$ m/s, the trend mirrors that of the final zombie fraction, albeit with noticeable differences observed for lower values of $v_z^{max}$. This occurs because a small portion of the realizations fail to achieve total conversion. This is primarily caused by multiple zombies colliding among themselves while pursuing a single human agent, preventing them from catching it. Furthermore, in instances where total conversion does not occur, the final fraction tends to be significantly high, usually around $\phi_z \sim \frac{N_h}{N_h+1}$. This explains the disparities between panels A and B of Fig. \ref{fig:success_rate} when the maximum zombie speed is lower than the human speed. The video of a specific realization for this scenario is available in the Supplemental Material \cite{SMV5}, with $v_z^{max}=3.8$ m/s being less than $v_h^{max}$.

\subsubsection{Total Conversion Times}

\begin{figure}
    \centering
    \begin{tikzpicture}
             \draw (0, 0) node[inner sep=0] {\includegraphics[width=\linewidth]{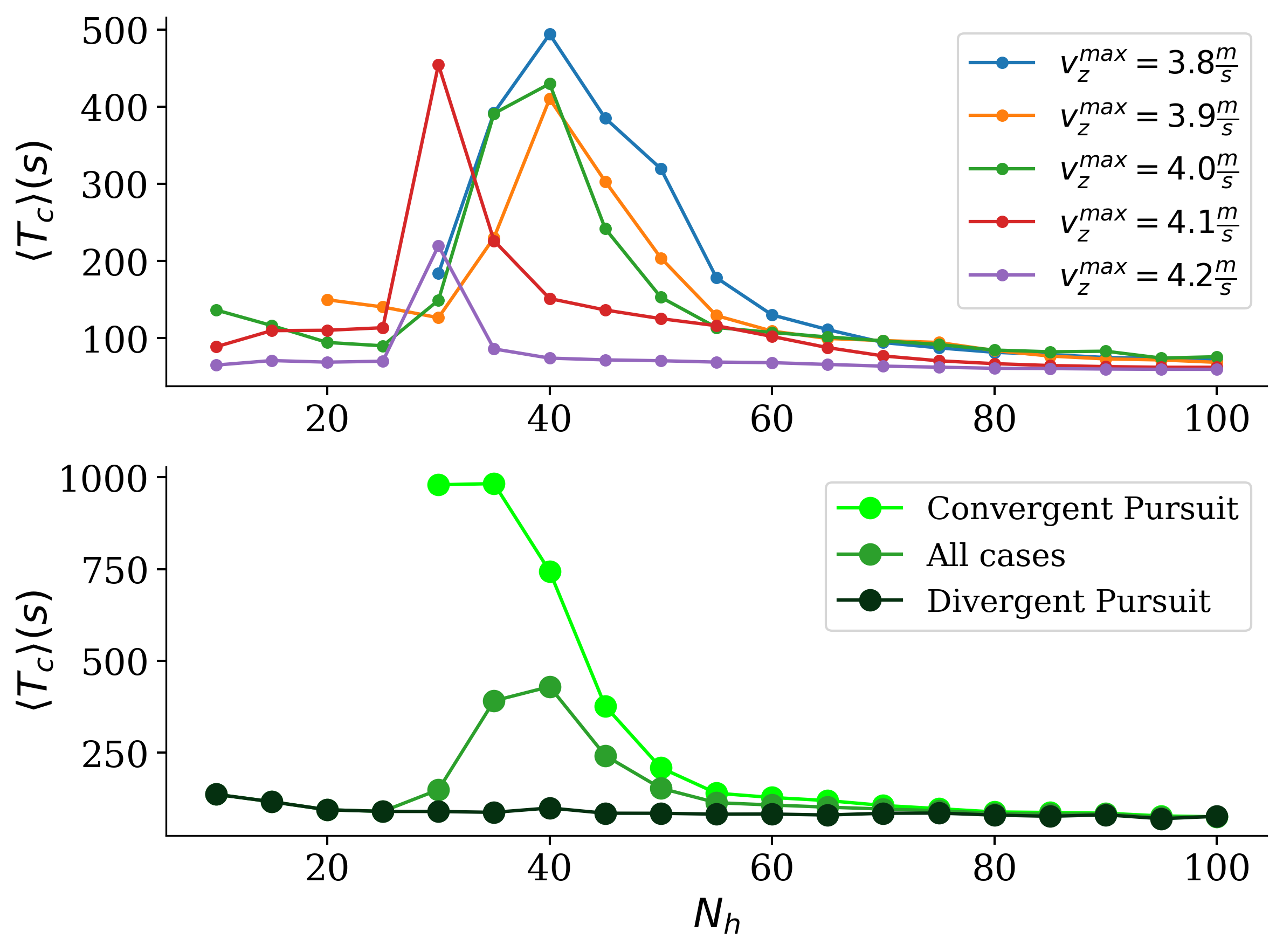}};
            \draw (-2.9,3.0) node[text=black, font=\sffamily] {\large A};
            \draw (-2.9,0.1) node[text=black, font=\sffamily] {\large B};
         \end{tikzpicture}
    \caption{Average Total Conversion Time. A: Across different configurations of $v_z^{max}$. B: Categorized by the type of pursuit for the scenario where $v_z^{max}=v_h^{max}=4$ m/s.}
    \label{fig:conversion_time}
\end{figure}

Now, we turn our attention to the study of the time needed to achieve total conversion. As Fig.~\ref{fig:success_rate}B clearly shows, total conversion is not always attained, being more probable at higher values of $N_h$ and $v_z^{max}$. Hence, the pertinent simulations within this context constitute a subset of those discussed in the preceding section. With this in mind, Fig.~\ref{fig:conversion_time}A illustrates the average total conversion time as a function of the initial number of human agents and for different maximum zombie speeds. Intriguingly, all curves exhibit a distinct maximum for all zombie speeds. In scenarios where zombie speeds exceed human speed, the peak conversion time occurs near $N_h \sim 30$, while for zombie speeds lower than or equal to human speed, the peak is around $N_h \sim 40$.

This suggests that achieving complete contagion occurs relatively quickly, irrespective of whether the initial number of humans is small or large. With fewer humans, they are less easily captured, yet the total conversion time remains low due to the smaller count of agents requiring conversion. Conversely, scenarios with a high number of humans result in easier capture (as density reduces their speed). At intermediate human numbers, a balance between these effects emerges: they are not easily captured, yet their adequate numbers contribute to slowing down the process.

Additional insight is presented in Fig.~\ref{fig:conversion_time}B, where scenarios of convergent and divergent pursuit are delineated for the case of zombie speed equal to human speed ($v_z^{max}=v_h^{max}$). Convergent pursuit results in total conversion only for $N_h \geq 30$, and, as expected, it represents a slow process that results in prolonged conversion times for intermediate values of $N_h$. In contrast,  divergent pursuit serves as the sole mechanism for achieving total conversion at lower $N_h$ values, occurring more rapidly. The peak observed in total conversion times distinctly arises from averaging the swift conversions in divergent pursuit with the slower, diminishing conversion times linked to convergent pursuit.

\subsubsection{Mean Population Velocities}

Previous studies have suggested a shift in the system's behavior at intermediate $N_h$ values. To delve deeper into this phenomenon, we will now investigate the system's mean speed.

The determination of mean velocity becomes less straightforward when humans and zombies possess different maximum speeds. In cases where $v_z^{max} < v_h^{max}$, the mean speed at low $N_h$ takes an indefinitely long time to stabilize. Conversely, in scenarios where $v_z^{max} > v_h^{max}$, simulations with high $N_h$ values typically result in total conversions before the mean speed can stabilize.

\begin{figure}
    \centering
      \begin{tikzpicture}
             \draw (0, 0) node[inner sep=0] {\includegraphics[width=\linewidth]{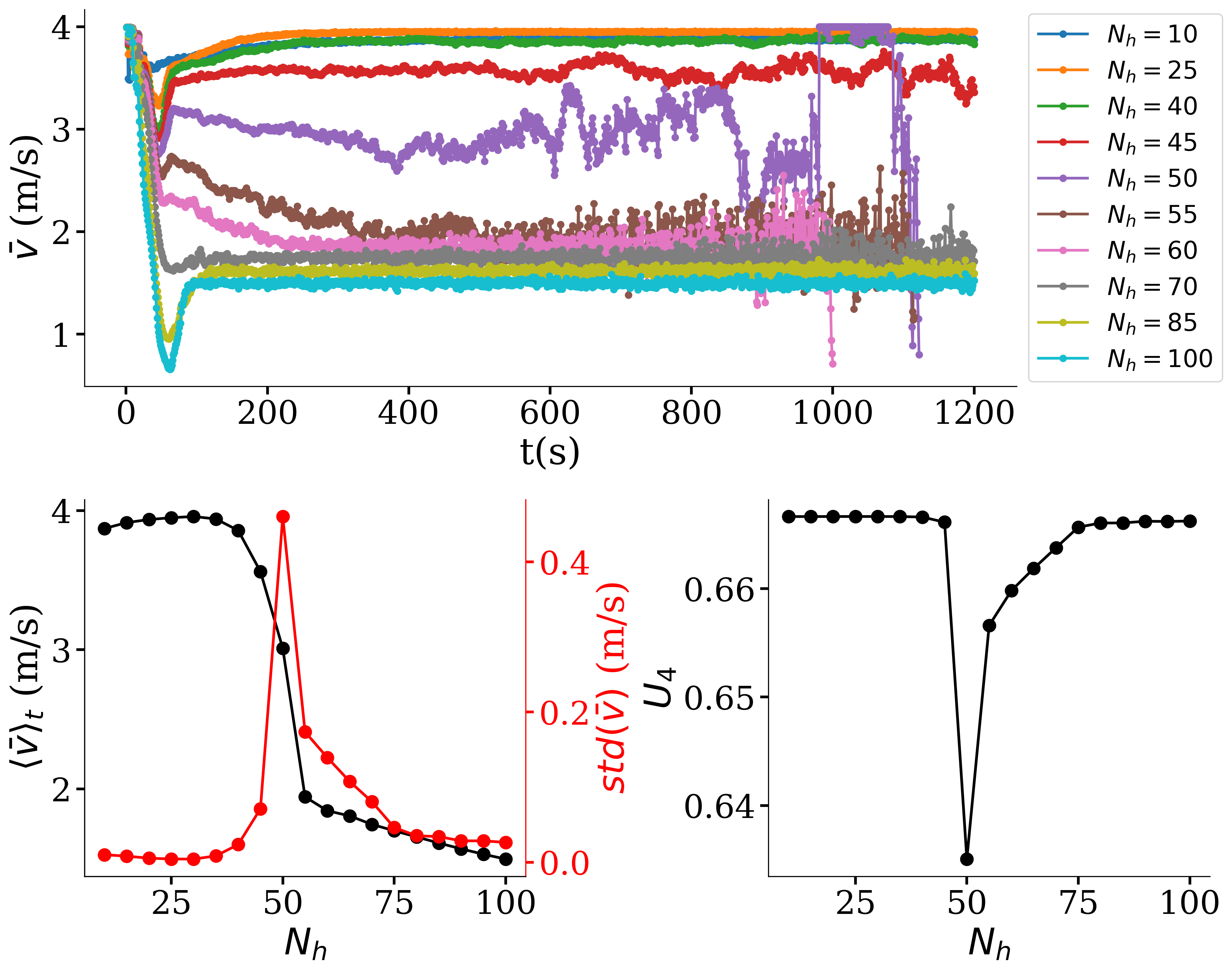}};
            \draw (-4.2,3.3) node[text=black, font=\sffamily] {\large A};
            \draw (-4.2,-0.2) node[text=black, font=\sffamily] {\large B};
            \draw (0.70,-0.2) node[text=black, font=\sffamily] {\large C};
         \end{tikzpicture}
    \caption{Population velocities for the scenario $v_z^{max} = v_h^{max} = 4$ m/s with various $N_h$ values. A: Evolution of population velocity averaged across all realizations. B: Mean population velocities after $t=250$ s, accompanied by the corresponding standard deviation. C: Binder parameter computed using the same data as in panel B. }
    \label{fig:speed_transition}
\end{figure}
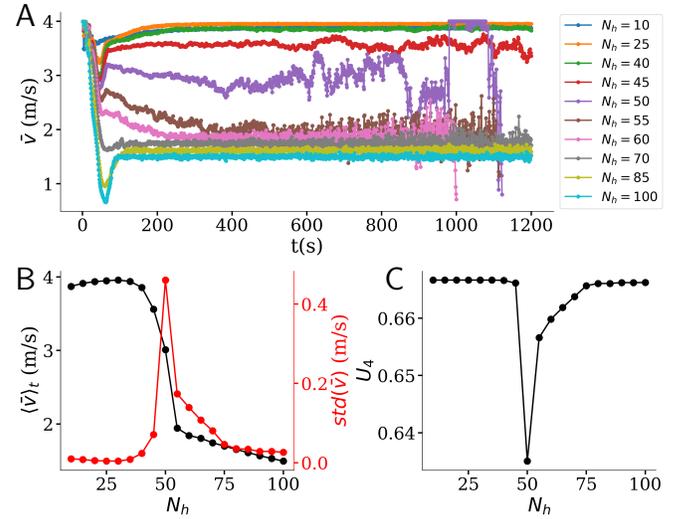

Stationary mean population speeds are evident solely when humans and zombies share equal maximum speeds ($v_z^{max} = v_h^{max}$), as depicted in Fig. \ref{fig:speed_transition}A. This figure also demonstrates that the stationary speed for low $N_h$ is close to the maximum ($\bar{v} \approx v_z^{max} \approx v_h^{max} \approx 4$ m/s), while for high $N_h$, the mean speed falls below $2$ m/s. At intermediate $N_h$ values, there is a noticeable transition in mean speed between these two groups.

This transition becomes clearer when examining the temporal mean in the stationary regime of mean speed (considered after $t=250$ s), as shown in Fig. \ref{fig:speed_transition}B. The standard deviations of the mean speed peak at $N_h=50$, suggesting a potential phase transition at the critical value $N_{hc}\approx 50$. In Fig. \ref{fig:speed_transition}C, the fourth-order Binder cumulant $U_4 = 1 - \frac{\left\langle\bar{v}^4\right\rangle_t}{3\left\langle\bar{v}^2\right\rangle_t^2}$ is computed for the same data, exhibiting a distinct minimum at $N_{hc}$, indicative of a first-order transition \cite{puzzo2022short}.

\section{Conclusions}
\label{sec:conc}

Employing an operational model of pedestrian dynamics \cite{Bagli2011}, known for accurately replicating the decrease in average speeds based on density (fundamental diagram of pedestrian traffic), we enhanced it by integrating evasion and pursuit mechanisms contingent upon the agent's type. The prey agents (humans) transition to the predator state (zombies) upon being captured by the predator.

This novel system not only acts as a test scenario for implementing the collision avoidance paradigm, where alterations are made solely to the agents' desired velocity, but it also exhibits a non-trivial emergent dynamic.

In basic scenarios involving two or three agents, it was observed that within a confined space, a predator can consistently catch its prey, even when they possess identical maximum speeds. However, when two predators pursue the same prey, their interference diminishes their speed, preventing them from ever capturing the sole prey.

The mutual collision mechanism can be disrupted when multiple preys exist, and each predator pursues a different prey. This scenario, termed 'divergent pursuit', becomes a crucial factor in achieving total conversion in environments characterized by lower population density.

In scenarios involving numerous agents, we regard the global density ($N_h$, the initial number of humans) as the controlling parameter. Both the final zombie fraction and the average times for total conversion manifest a distinct behavioral change at intermediate $N_h$ values. In the case of the final zombie fraction, it abruptly transitions from $\phi_z^{final}=0.5$ to $\phi_z^{final}=1$, and for the total conversion times, a peak is observed at values of $N_h$ that depend on the maximum zombie speed.

Remarkably, when considering the average population velocity as an order parameter in scenarios where the maximum speeds of humans and zombies are equal, the results present evidence of a first-order phase transition occurring at the critical value of $N_{hc}\approx 50$.

The proposed human-zombie scenario could be expanded into more complex versions by including several probabilistic elements: the chance that a human can eliminate a zombie, the likelihood of a human being immune, and the possibility of a zombie being cured and reverting to a human state. These additions would introduce new dynamics and complexities to the system.

\section{Acknowledgments}
    This work was funded by project PICT 2019-00511 (Agencia Nacional de Promoci\'on Cient\'ifica y Tecnol\'ogica, Argentina). 

\bibliographystyle{apsrev}
\bibliography{Pedestrian_vs_Zombies.bib}% Produces the 

\begin{thebibliography}{19}
\expandafter\ifx\csname natexlab\endcsname\relax\def\natexlab#1{#1}\fi
\expandafter\ifx\csname bibnamefont\endcsname\relax
  \def\bibnamefont#1{#1}\fi
\expandafter\ifx\csname bibfnamefont\endcsname\relax
  \def\bibfnamefont#1{#1}\fi
\expandafter\ifx\csname citenamefont\endcsname\relax
  \def\citenamefont#1{#1}\fi
\expandafter\ifx\csname url\endcsname\relax
  \def\url#1{\texttt{#1}}\fi
\expandafter\ifx\csname urlprefix\endcsname\relax\def\urlprefix{URL }\fi
\providecommand{\bibinfo}[2]{#2}
\providecommand{\eprint}[2][]{\url{#2}}

\bibitem[{\citenamefont{Helbing and Molnár}(1995)}]{Helb1995}
\bibinfo{author}{\bibfnamefont{D.}~\bibnamefont{Helbing}} \bibnamefont{and} \bibinfo{author}{\bibfnamefont{P.}~\bibnamefont{Molnár}}, \bibinfo{journal}{Physical Review E} \textbf{\bibinfo{volume}{51}}, \bibinfo{pages}{4282} (\bibinfo{year}{1995}).

\bibitem[{\citenamefont{Helbing et~al.}(2000)\citenamefont{Helbing, Farkas, and Vicsek}}]{Helb2000}
\bibinfo{author}{\bibfnamefont{D.}~\bibnamefont{Helbing}}, \bibinfo{author}{\bibfnamefont{I.}~\bibnamefont{Farkas}}, \bibnamefont{and} \bibinfo{author}{\bibfnamefont{T.}~\bibnamefont{Vicsek}}, \bibinfo{journal}{Nature} \textbf{\bibinfo{volume}{407}}, \bibinfo{pages}{487} (\bibinfo{year}{2000}).

\bibitem[{\citenamefont{Karamouzas et~al.}(2009)\citenamefont{Karamouzas, Heil, Van~Beek, and Overmars}}]{karamouzas2009predictive}
\bibinfo{author}{\bibfnamefont{I.}~\bibnamefont{Karamouzas}}, \bibinfo{author}{\bibfnamefont{P.}~\bibnamefont{Heil}}, \bibinfo{author}{\bibfnamefont{P.}~\bibnamefont{Van~Beek}}, \bibnamefont{and} \bibinfo{author}{\bibfnamefont{M.~H.} \bibnamefont{Overmars}}, in \emph{\bibinfo{booktitle}{Motion in Games: Second International Workshop, MIG 2009, Zeist, The Netherlands, November 21-24, 2009. Proceedings 2}} (\bibinfo{organization}{Springer}, \bibinfo{year}{2009}), pp. \bibinfo{pages}{41--52}.

\bibitem[{\citenamefont{Wang et~al.}(2015)\citenamefont{Wang, Chen, Dong, Zhou, and Ning}}]{wang2015new}
\bibinfo{author}{\bibfnamefont{Q.-L.} \bibnamefont{Wang}}, \bibinfo{author}{\bibfnamefont{Y.}~\bibnamefont{Chen}}, \bibinfo{author}{\bibfnamefont{H.-R.} \bibnamefont{Dong}}, \bibinfo{author}{\bibfnamefont{M.}~\bibnamefont{Zhou}}, \bibnamefont{and} \bibinfo{author}{\bibfnamefont{B.}~\bibnamefont{Ning}}, \bibinfo{journal}{Chinese Physics B} \textbf{\bibinfo{volume}{24}}, \bibinfo{pages}{038901} (\bibinfo{year}{2015}).

\bibitem[{\citenamefont{Martin and Parisi}(2020{\natexlab{a}})}]{martin2020pedestrian}
\bibinfo{author}{\bibfnamefont{R.}~\bibnamefont{Martin}} \bibnamefont{and} \bibinfo{author}{\bibfnamefont{D.}~\bibnamefont{Parisi}}, \bibinfo{journal}{Collective Dynamics} \textbf{\bibinfo{volume}{5}}, \bibinfo{pages}{324} (\bibinfo{year}{2020}{\natexlab{a}}).

\bibitem[{\citenamefont{Martin and Parisi}(2020{\natexlab{b}})}]{Martin2020}
\bibinfo{author}{\bibfnamefont{R.}~\bibnamefont{Martin}} \bibnamefont{and} \bibinfo{author}{\bibfnamefont{D.}~\bibnamefont{Parisi}}, \bibinfo{journal}{Collective Dynamics} \textbf{\bibinfo{volume}{5}} (\bibinfo{year}{2020}{\natexlab{b}}).

\bibitem[{\citenamefont{Martin and Parisi}(2020{\natexlab{c}})}]{martin2020data}
\bibinfo{author}{\bibfnamefont{R.~F.} \bibnamefont{Martin}} \bibnamefont{and} \bibinfo{author}{\bibfnamefont{D.~R.} \bibnamefont{Parisi}}, in \emph{\bibinfo{booktitle}{Traffic and Granular Flow 2019}} (\bibinfo{organization}{Springer}, \bibinfo{year}{2020}{\natexlab{c}}), pp. \bibinfo{pages}{205--210}.

\bibitem[{\citenamefont{Martin and Parisi}(2023)}]{martin2023anisotropic}
\bibinfo{author}{\bibfnamefont{R.~F.} \bibnamefont{Martin}} \bibnamefont{and} \bibinfo{author}{\bibfnamefont{D.~R.} \bibnamefont{Parisi}}, \bibinfo{journal}{Physica A: Statistical Mechanics and its Applications} p. \bibinfo{pages}{129414} (\bibinfo{year}{2023}).

\bibitem[{\citenamefont{Munz et~al.}(2009)\citenamefont{Munz, Hudea, Imad, and Smith}}]{munz2009zombies}
\bibinfo{author}{\bibfnamefont{P.}~\bibnamefont{Munz}}, \bibinfo{author}{\bibfnamefont{I.}~\bibnamefont{Hudea}}, \bibinfo{author}{\bibfnamefont{J.}~\bibnamefont{Imad}}, \bibnamefont{and} \bibinfo{author}{\bibfnamefont{R.~J.} \bibnamefont{Smith}}, \bibinfo{journal}{Infectious disease modelling research progress} \textbf{\bibinfo{volume}{4}}, \bibinfo{pages}{133} (\bibinfo{year}{2009}).

\bibitem[{\citenamefont{Alemi et~al.}(2015)\citenamefont{Alemi, Bierbaum, Myers, and Sethna}}]{alemi2015you}
\bibinfo{author}{\bibfnamefont{A.~A.} \bibnamefont{Alemi}}, \bibinfo{author}{\bibfnamefont{M.}~\bibnamefont{Bierbaum}}, \bibinfo{author}{\bibfnamefont{C.~R.} \bibnamefont{Myers}}, \bibnamefont{and} \bibinfo{author}{\bibfnamefont{J.~P.} \bibnamefont{Sethna}}, \bibinfo{journal}{Physical Review E} \textbf{\bibinfo{volume}{92}}, \bibinfo{pages}{052801} (\bibinfo{year}{2015}).

\bibitem[{\citenamefont{Bauer}(2019)}]{bauer2019mathematical}
\bibinfo{author}{\bibfnamefont{H.}~\bibnamefont{Bauer}}, \bibinfo{type}{Undergraduate thesis}, \bibinfo{school}{University of Lynchburg} (\bibinfo{year}{2019}), \urlprefix\url{https://digitalshowcase.lynchburg.edu/cgi/viewcontent.cgi?article=1122&context=utcp}.

\bibitem[{\citenamefont{Lib{\'a}l et~al.}(2023)\citenamefont{Lib{\'a}l, Forg{\'a}cs, N{\'e}da, Reichhardt, Hengartner, and Reichhardt}}]{libal2023transition}
\bibinfo{author}{\bibfnamefont{A.}~\bibnamefont{Lib{\'a}l}}, \bibinfo{author}{\bibfnamefont{P.}~\bibnamefont{Forg{\'a}cs}}, \bibinfo{author}{\bibfnamefont{{\'A}.}~\bibnamefont{N{\'e}da}}, \bibinfo{author}{\bibfnamefont{C.}~\bibnamefont{Reichhardt}}, \bibinfo{author}{\bibfnamefont{N.}~\bibnamefont{Hengartner}}, \bibnamefont{and} \bibinfo{author}{\bibfnamefont{C.}~\bibnamefont{Reichhardt}}, \bibinfo{journal}{Physical Review E} \textbf{\bibinfo{volume}{107}}, \bibinfo{pages}{024604} (\bibinfo{year}{2023}).

\bibitem[{\citenamefont{Baglietto and Parisi}(2011)}]{Bagli2011}
\bibinfo{author}{\bibfnamefont{G.}~\bibnamefont{Baglietto}} \bibnamefont{and} \bibinfo{author}{\bibfnamefont{D.~R.} \bibnamefont{Parisi}}, \bibinfo{journal}{Physical Review E} \textbf{\bibinfo{volume}{83}} (\bibinfo{year}{2011}).

\bibitem[{SMV({\natexlab{a}})}]{SMV1}
\bibinfo{note}{\url{https://youtu.be/ZCp7hJBvf14}}.

\bibitem[{SMV({\natexlab{b}})}]{SMV2}
\bibinfo{note}{\url{https://youtu.be/84liICIE0Ek}}.

\bibitem[{SMV({\natexlab{c}})}]{SMV3}
\bibinfo{note}{\url{https://youtu.be/YDJkbw_ayiQ}}.

\bibitem[{SMV({\natexlab{d}})}]{SMV4}
\bibinfo{note}{\url{https://youtu.be/g8WlDL1h65A}}.

\bibitem[{SMV({\natexlab{e}})}]{SMV5}
\bibinfo{note}{\url{https://youtu.be/yIMXLRIe1mY}}.

\bibitem[{\citenamefont{Puzzo et~al.}(2022)\citenamefont{Puzzo, Loscar, De~Virgiliis, and Grigera}}]{puzzo2022short}
\bibinfo{author}{\bibfnamefont{M.~L.~R.} \bibnamefont{Puzzo}}, \bibinfo{author}{\bibfnamefont{E.~S.} \bibnamefont{Loscar}}, \bibinfo{author}{\bibfnamefont{A.}~\bibnamefont{De~Virgiliis}}, \bibnamefont{and} \bibinfo{author}{\bibfnamefont{T.~S.} \bibnamefont{Grigera}}, \bibinfo{journal}{Journal of Physics: Condensed Matter} \textbf{\bibinfo{volume}{34}}, \bibinfo{pages}{314001} (\bibinfo{year}{2022}).

\end{thebibliography}

\end{document}